\documentclass[final]{article}

\usepackage{scicite}
\usepackage{times}
\usepackage{graphicx}

\newenvironment{sciabstract}{%
\begin{quote} \bf}
{\end{quote}}

\newcounter{lastnote}
\newenvironment{scilastnote}{%
\setcounter{lastnote}{\value{enumiv}}%
\addtocounter{lastnote}{+1}%
\begin{list}%
{\arabic{lastnote}.}
{\setlength{\leftmargin}{.22in}}
{\setlength{\labelsep}{.5em}}}
{\end{list}}

\begin{document} 
%\begin{singlespace}
\baselineskip24pt
%
% The long author list broke the formatting of the maketitle command.  
% Typed the information directly in LaTeX following Science's format.
%
\begin{center}
\LARGE{Probing Cold Dense Nuclear Matter}
\end{center}
\large{\begin{center}
R.~Subedi,$^{1}$  R.~Shneor,$^{2}$ P.~Monaghan,$^{3}$ B.~D.~Anderson,$^{1}$ K.~Aniol,$^{4}$  J.~Annand,$^{5}$ 
J.~Arrington,$^{6}$ H.~Benaoum,$^{7,8}$ F.~Benmokhtar,$^{9}$ W.~Bertozzi,$^{3}$ W.~Boeglin,$^{10}$ 
J.-P.~Chen,$^{11}$ Seonho~Choi,$^{12}$ E.~Cisbani,$^{13}$ B.~Craver,$^{14}$  
S.~Frullani,$^{13}$ F.~Garibaldi,$^{13}$ S.~Gilad,$^{3}$ R.~Gilman,$^{11,15}$ 
O.~Glamazdin,$^{16}$ J.-O.~Hansen,$^{11}$ D.~W.~Higinbotham,$^{11\ast}$ 
T.~Holmstrom,$^{17}$ 
H.~Ibrahim,$^{18}$ R.~Igarashi,$^{19}$ C.W.~de~Jager,$^{11}$ E.~Jans,$^{20}$  X.~Jiang,$^{15}$ L.J.~Kaufman,$^{9,22}$ 
A.~Kelleher,$^{17}$ A.~Kolarkar,$^{23}$ G.~Kumbartzki,$^{15}$ J.~J.~LeRose,$^{11}$ 
R.~Lindgren,$^{14}$ N.~Liyanage,$^{14}$ D.~J.~Margaziotis,$^{4}$ P.~Markowitz,$^{10}$ S.~Marrone,$^{24}$ 
M.~Mazouz,$^{25}$ D.~Meekins,$^{11}$ R.~Michaels,$^{11}$ B.~Moffit,$^{17}$ C.~F.~Perdrisat,$^{17}$ 
E.~Piasetzky,$^{2}$ M.~Potokar,$^{26}$ V.~Punjabi,$^{27}$ Y.~Qiang,$^{3}$ J.~Reinhold,$^{10}$
G.~Ron,$^{2}$ G.~Rosner,$^{28}$ A.~Saha,$^{11}$ B.~Sawatzky,$^{14,29}$ A.~Shahinyan,$^{30}$ S.~\v{S}irca,$^{26,31}$ 
K.~Slifer,$^{14}$ P.~Solvignon,$^{29}$ V.~Sulkosky,$^{17}$  
G.~M.~Urciuoli,$^{13}$ E.~Voutier,$^{25}$ J.~W.~Watson,$^{1}$  L.B.~Weinstein,$^{18}$  
B.~Wojtsekhowski,$^{11}$ S.~Wood,$^{11}$ X.-C.~Zheng,$^{3,6,14}$ and L.~Zhu$^{32}$ 
\end{center}

\begin{center}
    \normalsize{$^{1}$Kent State University, Kent State, OH 44242, USA}\\
    \normalsize{$^{2}$Tel Aviv University, Tel Aviv 69978, Israel}\\
    \normalsize{$^{3}$Massachusetts Institute of Technology, Cambridge, MA 02139, USA}\\
    \normalsize{$^{4}$California State University Los Angeles, Los Angeles, CA 90032, USA}\\
    \normalsize{$^{5}$University of Glasgow, Glasgow G12 8QQ, Scotland, UK}\\
    \normalsize{$^{6}$Argonne National Laboratory, Argonne, IL 60439, USA}\\
    \normalsize{$^{7}$Syracuse University, Syracuse, NY 13244, USA}\\
    \normalsize{$^{8}$Prince Mohammad University, Al-Khobar 31952, Saudi Arabia}\\
    \normalsize{$^{9}$University of Maryland, College Park. MD 20742, USA}\\
    \normalsize{$^{10}$Florida International University, Miami, FL 33199, USA}\\
    \normalsize{$^{11}$Thomas Jefferson National Accelerator Facility, Newport News, VA 23606, USA}\\
    \normalsize{$^{12}$Seoul National University, Seoul 151-747, Korea}\\
    \normalsize{$^{13}$INFN, Sezione di Roma, I-00185 Rome, Italy}\\
    \normalsize{$^{14}$University of Virginia, Charlottesville, VA 22904, USA}\\
    \normalsize{$^{15}$Rutgers, The State University of New Jersey, Piscataway, NJ 08855, USA}\\
    \normalsize{$^{16}$Kharkov Institute of Physics and Technology, Kharkov 310108, Ukraine}\\
    \normalsize{$^{17}$College of William and Mary, Williamsburg, VA 23187, USA}\\
    \normalsize{$^{18}$Old Dominion University, Norfolk, VA 23508, USA}\\
    \normalsize{$^{19}$University of Saskatchewan, Saskatoon, Saskatchewan, Canada S7N 5E2}\\
    \normalsize{$^{20}$Nationaal Instituut voor Subatomaire Fysica, Amsterdam, The Netherlands}\\
    \normalsize{$^{22}$University of Massachusetts Amherst, Amherst, MA 01003, USA}\\
    \normalsize{$^{23}$University of Kentucky, Lexington, KY 40506, USA}\\
    \normalsize{$^{24}$Dipartimento di Fisica and INFN sez. Bari, Bari, Italy}\\
    \normalsize{$^{25}$Laboratoire de Physique Subatomique et de Cosmologie, 38026 Grenoble, France}\\
    \normalsize{$^{26}$Institute ``Jo\v{z}ef Stefan'', 1000 Ljubljana, Slovenia}\\
    \normalsize{$^{27}$Norfolk State University, Norfolk, VA 23504, USA}\\
    \normalsize{$^{28}$University of Glasgow, Glasgow G12 8QQ, Scotland, UK}\\
    \normalsize{$^{29}$Temple University, Philadelphia, PA 19122, USA}\\
    \normalsize{$^{30}$Yerevan Physics Institute, Yerevan 375036, Armenia}\\
    \normalsize{$^{31}$Dept. of Physics, University of Ljubljana, 1000 Ljubljana, Slovenia}\\
    \normalsize{$^{32}$University of Illinois at Urbana-Champaign, Urbana, IL 61801, USA} \\
    \vspace{5mm}
    \normalsize{$^\ast$To whom correspondence should be addressed; E-mail: doug@jlab.org.}
\end{center}
}
%\end{singlespace}

\clearpage

\begin{sciabstract}
The protons and neutrons in a nucleus can
form strongly correlated nucleon pairs.  
Scattering experiments,
where a proton is knocked-out of the nucleus 
with high momentum transfer and high missing momentum,
show that in
$^{12}$C the neutron-proton pairs are nearly twenty times as prevalent 
as proton-proton pairs and, by inference, neutron-neutron pairs. 
This difference between the types of pairs is due to the nature of the strong force  
and has implications for understanding cold dense nuclear 
systems such as neutron stars.
\end{sciabstract}

\section*{Introduction}

Nuclei are composed of bound protons and neutrons, referred to collectively as 
nucleons (the standard notation is p, n, and N, respectively). 
A standard model of the nucleus since the 1950s has been the nuclear shell model, 
where neutrons and protons move independently in well-defined quantum orbits in the average nuclear 
field created by their mutual attractive interactions. 
In the 1980s and 1990s, proton removal experiments using electron beams with energies of several hundred MeV
showed that only 60-70\% of the protons participate in this type of independent-particle 
motion in nuclear valence states~\cite{Lapikas:1993,Kelly:1996}.  
At the time, it was assumed that this low occupancy was caused by correlated pairs of 
nucleons within the nucleus. Indeed, the existence of nucleon pairs that 
are correlated at distances of several femtometers, known as long-range correlations,  
has been established~\cite{Dickhoff:2004xx},
but these accounted for less than half of the predicted correlated nucleon pairs.  
Recent high momentum transfer 
measurements~\cite{Egiyan:2003vg,Egiyan:2005hs,Niyazov:2003zr,Benmokhtar:2004fs,Aclander:1999fd,Tang:2002ww,Malki:2000gh,Piasetzky:2006ai,Shneor:2007tu} 
have shown that nucleons in nuclear ground states can form pairs 
with large relative momentum and small center-of-mass (CM) momentum
due to the short-range, scalar and tensor, components of the
nucleon-nucleon interaction.
These pairs
are referred to as short-range correlated (SRC) pairs.  
The study of these SRC pairs allows 
access to cold dense nuclear matter,
such as that found in a neutron star.

Experimentally, a high-momentum probe can knock a proton out of a nucleus, leaving the rest 
of the system nearly unaffected.  If, on the other hand, the proton being struck is 
part of a SRC pair, the  high relative momentum in the pair would cause 
the correlated nucleon to recoil and be ejected as well (Fig.~1).  
High-momentum knock-out by both high-energy
protons~\cite{Aclander:1999fd,Tang:2002ww,Malki:2000gh} and high-energy electrons~\cite{Shneor:2007tu} 
has shown, for kinematics far from particle production resonances,
that when a proton with high missing momentum is removed from the $^{12}$C nucleus, the momentum is 
predominantly balanced by a single recoiling nucleon. 
This is consistent with the theoretical description that
large nucleon momenta in the nucleus are predominantly caused by SRC pairing~\cite{Frankfurt:1981mk}.
This effect has also been shown using inclusive (e,e') data~\cite{Egiyan:2003vg,Egiyan:2005hs,Frankfurt:1993sp}, 
though that type of measurement is not sensitive to the type of SRC pair.
Here, we identify the relative abundance of p-n and p-p SRC pairs in $^{12}$C nuclei. 

\begin{figure}[htb]
\includegraphics[width=\linewidth]{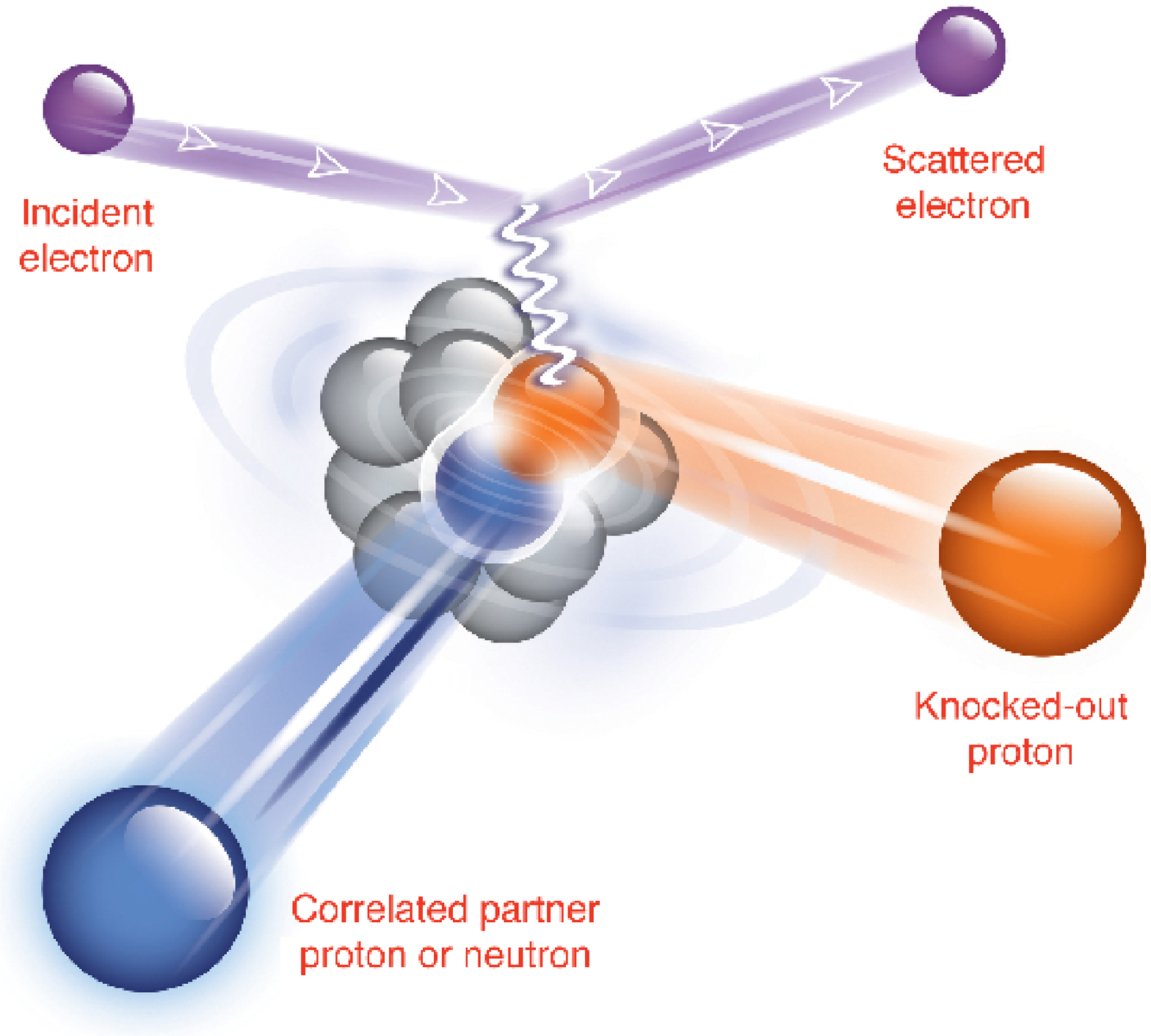}
\caption{
Illustration of the $^{12}$C(e,e'pN) reaction.  The incident electron beam couples 
to a nucleon-nucleon pair via
a virtual photon.  In the final state, the scattered electron is detected along 
with the two nucleons that are 
ejected from
the nucleus.  Typical nuclear density is about 0.16~nucleons/fm$^{3}$ while 
for pairs the local density is approximately 5 times larger.  
}
\label{reaction}
\end{figure}

\section*{Equipment}

We performed our experiment
in Hall A of the Thomas Jefferson National Accelerator Facility (JLab) 
using an incident electron beam of 4.627~GeV with a beam current between 5 and 40~$\mu$A.  
The beam was incident on a 0.25~mm thick pure 
$^{12}$C sheet rotated 70$^{\circ}$ to the beam line to minimize the material through which the 
recoiling protons passed.  We used the 
two Hall A high-resolution spectrometers 
(HRS)~\cite{Alcorn:2004sb} to define proton-knockout events, $^{12}$C(e,e'p).  
The left HRS detected scattered electrons at a central scattering angle (momentum) of 
19.5$^{\circ}$ (3.724 GeV/c).  These values correspond to the quasi-free knockout of a 
single proton with 
transferred three-momentum q = 1.65~GeV/c, transferred energy $\omega$ = 0.865 GeV, 
Q$^2$ = q$^2 - (\omega/c)^2$ = 2~(GeV/c)$^2$, 
and Bjorken scaling parameter $x_B$ = Q$^2$/2m$\omega$ = 1.2, where m is the mass of the proton. 
The right HRS  detected knocked-out protons at three different values for the central angle (momentum): 
40.1$^{\circ}$ (1.45 GeV/c), 35.8$^{\circ}$ (1.42 GeV/c), and 32.0$^{\circ}$ (1.36 GeV/c).  
These kinematic settings covered (e,e'p) 
missing momenta, which is the momentum of the undetected particles, in 
the range of 300-600~MeV/c with overlap between the different settings.  
For highly correlated pairs, the 
missing momentum of the (e,e'p) reaction is balanced almost 
entirely by a single recoiling nucleon; whereas for
a typical uncorrelated (e,e'p) event, the missing momentum 
is balanced by the sum of many recoiling nucleons.   
In a partonic picture, $x_B$ is the fraction of the nucleon momentum carried 
by the struck quark.  Hence, when $x_B > 1$, the struck quark has more momentum 
than the entire nucleon, which points to nucleon correlation.  
To detect correlated recoiling protons, a large acceptance spectrometer (BigBite) was placed 
at an angle of 99$^{\circ}$ w.r.t. the beam direction and 1.1~m from the target.
To detect correlated recoiling neutrons, a neutron array was placed directly
behind the BigBite spectrometer at a distance of 6~m from the target.  Details of these
custom proton and neutron detectors can be found in the supporting on-line materials.

\section*{Analysis and Results}

The electronics for the experiment were set up such that for every $^{12}$C(e,e'p) event
in the HRS spectrometers, we read out the BigBite and the neutron-detector electronics;
thus, we could determine the 
$^{12}$C(e,e'pp)/$^{12}$C(e,e'p) and the $^{12}$C(e,e'pn)/$^{12}$C(e,e'p) ratios. 
For the ratio $^{12}$C(e,e'pp)/$^{12}$C(e,e'p), we found
that $9.5 \pm 2$\% of the (e,e'p) events had an 
associated recoiling proton as reported in~\cite{Shneor:2007tu}.  
Taking into account the finite acceptance of the neutron
detector, using the same procedure as was done for the proton detector~\cite{Shneor:2007tu},
and the neutron detection efficency, we found that $96 \pm 22$\% of the (e,e'p) events with a missing
momentum above 300~MeV/c had a
recoiling neutron.   This result agrees with a hadron beam measurement of (p,2pn)/(p,2p) 
in which $92 \pm 18$\% of the (p,2p) events with a missing momentum above the Fermi momentum of 275~MeV/c 
were found to have a single recoiling neutron carrying the momentum~\cite{Piasetzky:2006ai}. 

Since we collected the recoiling proton $^{12}$C(e,e'pp) and neutron $^{12}$C(e,e'pn) 
data simultaneously with detection systems covering 
nearly identical solid angles, we could also directly determine the ratio of $^{12}$C(e,e'pn)/$^{12}$C(e,e'pp). 
In this scheme, many of the systematic factors needed to compare the rates of the $^{12}$C(e,e'pn) 
and $^{12}$C(e,e'pp) reactions canceled out.  
Correcting only for detector efficiencies, we determined that this ratio was 8.1$\pm$2.2.
To estimate the effect of final-state interactions (i.e., reactions that happen after the initial scattering),
we assumed that the attenuation of the recoiling protons 
and neutrons were almost equal.  In this case, the only correction related 
to final-state interactions of the measured $^{12}$C(e,e'pn) to $^{12}$C(e,e'pp) ratio is due to single charge exchange.  
Since the measured (e,e'pn) rate is about an order of magnitude larger than the (e,e'pp) rate, (e,e'pn) 
reactions followed by single charge exchange (and hence detected as (e,e'pp)) dominate and 
reduced the measured $^{12}$C(e,e'pn)/$^{12}$C(e,e'pp) ratio. 
Using the Glauber approximation~\cite{Mardor:1992sb}, we estimated this effect 
was 11\%. 
Taking this into account, the corrected experimental ratio for $^{12}$C(e,e'pn)/$^{12}$C(e,e'pp) 
is 9.0$\pm$2.5.  

To deduce the ratio of p-n to p-p SRC pairs 
in the ground state of $^{12}$C,  
we use the measured $^{12}$C(e,e'pn) to $^{12}$C(e,e'pp) ratio.
Because we used (e,e'p)  
events to search for SRC nucleon pairs, the probability of detecting p-p pairs 
was twice that of p-n pairs;
thus, we conclude that 
the ratio of p-n/p-p pairs in the $^{12}$C ground state is $18\pm5$ as shown in Fig.~2.
To get a comprehensive picture of  the structure of $^{12}$C, we 
combined the pair faction results with the inclusive $^{12}$C(e,e') 
measurements~\cite{Egiyan:2003vg,Egiyan:2005hs,Frankfurt:1993sp} where it was found
that approximately 20\% of the nucleons in $^{12}$C form SRC pairs: a result
consistent with the the depletion seen in the spectroscopy experiments~\cite{Lapikas:1993,Kelly:1996}.
As shown in Fig.~3, the combined results indicate 80\% of the
nucleons in the $^{12}$C nucleus acted independent 
or as described within the shell model, whereas for the 20\% of correlated pairs,
%long-range correlated nucleons, 
90$\pm$10\% were in the form of p-n SRC pairs, 5$\pm$1.5\%
were in the form of p-p SRC pairs, and by 
isospin symmetry infer that 5$\pm$1.5\% were in the form of SRC n-n pairs.
The dominance of the p-n over p-p SRC pairs is a 
clear consequence of the nucleon-nucleon tensor force.
Calculations of this effect~\cite{rocco2007,Sargsian:2005ru}
indicate that it is robust and
does not depend on the exact parameterization of the nucleon-nucleon force, the type of the nucleus, 
or the exact ground-state wave-function used to describe the nucleons.

\begin{figure}[htb]
\includegraphics[angle=-90,width=\linewidth]{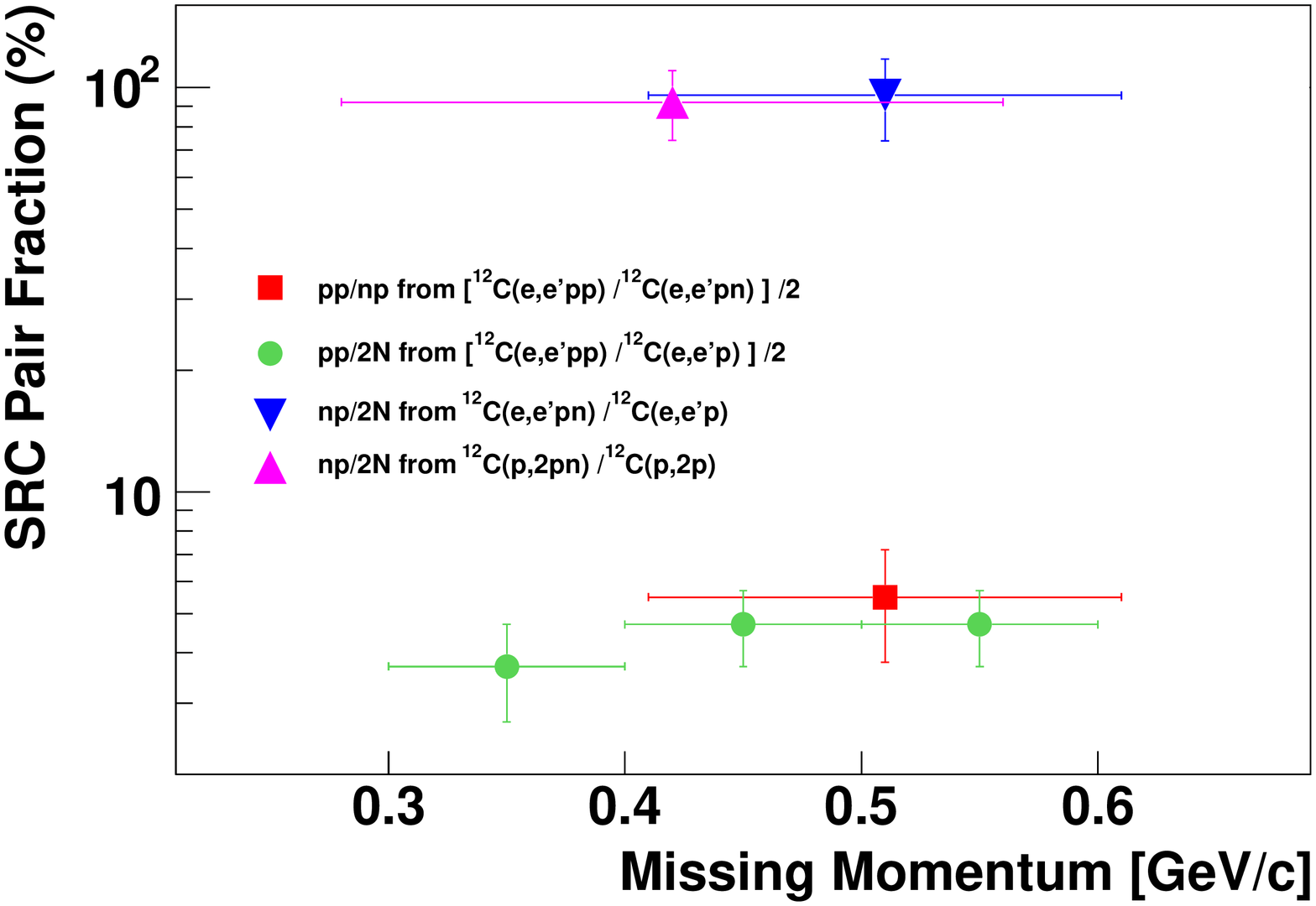}
\caption{The fractions of correlated pair combinations in carbon as obtained from
the (e,e'pp) and (e,e'pn) reactions, as well as from previous (p,2pn) data.  The results
and references are listed in Table~1.}
\label{results}
\end{figure}

\begin{figure}[htb]
\includegraphics[width=\linewidth]{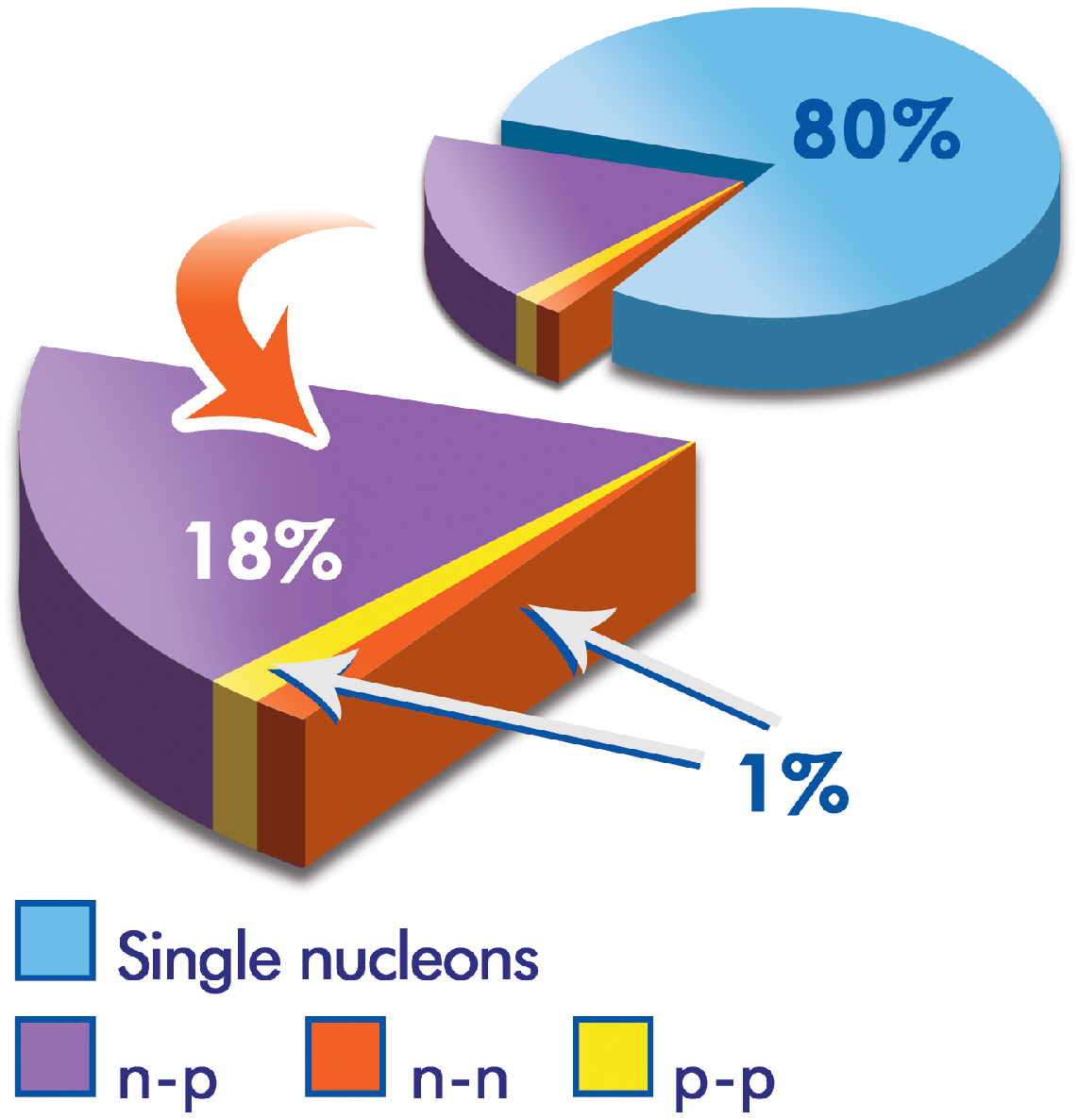}
\caption{
The average fraction of nucleons in the various initial state configurations of $^{12}$C.
}
\label{pie}
\end{figure}

If neutron stars consisted of neutrons only, the relatively
weak n-n short-range interaction would mean that they could be reasonably
well approximated as an ideal Fermi gas,  with only perturbative
corrections. However, theoretical analysis of neutrino cooling data
indicates that neutron stars contain
about 5-10\% protons and electrons in the first central layers~\cite{Lattimer:2004,Baym:1995,Baym:2002}.
The strong p-n short-range interaction reported here suggests that
momentum distribution for the protons and for the neutrons
in neutron stars will be substantially different from that characteristic of an ideal
Fermi gas.  A theoretical calculation that takes into account the p-n
correlation effect at relevant neutron star densities and realistic proton
concentration shows the correlation effect on the momentum distribution of
the protons and the neutrons~\cite{Frick:2004th}.  We therefore
speculate that the small concentration of protons inside neutron
stars might have a disproportionately large effect that needs to be
addressed in realistic descriptions of neutron stars.

\clearpage

\section*{Supporting Materials}

\begin{figure}[htb]
  \centering
  \includegraphics[width=\linewidth,angle=0]{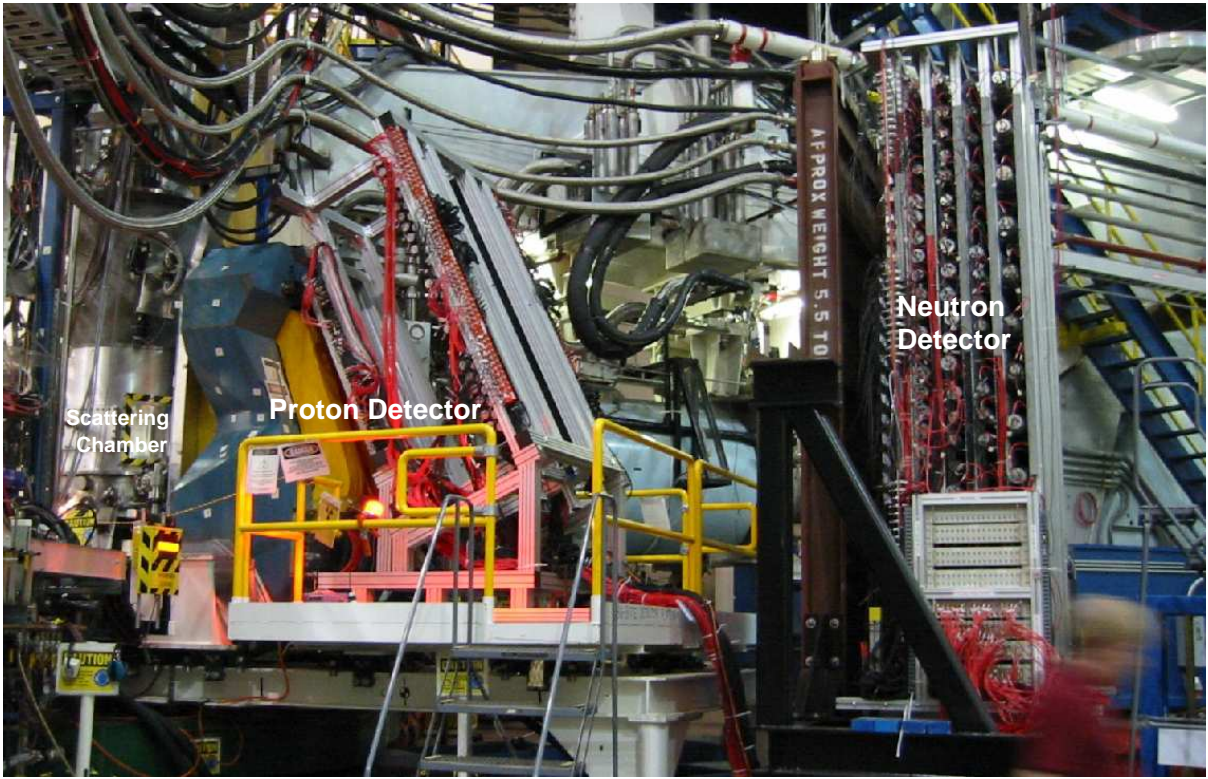}  
  \caption{Photograph of the proton and neutron detectors in experimental Hall~A.}  
  \label{bbnd-fig}
\end{figure}

The experiment was performed in Hall~A of the Continuous Electron Beam Accelerator Facility (CEBAF)
located at the Thomas Jefferson National Accelerator Facility in Newport News, VA~\cite{Leemann:2001}.  
Along with the standard Hall~A equipment~\cite{Alcorn:2004sb}, the experiment required a new proton spectrometer and neutron detector
as well as a new scattering chamber to accommodate the large out-of-plane acceptance
of these detectors (see Fig.~\ref{bbnd-fig}). 

\subsection*{Proton Spectrometer}

The proton spectrometer (BigBite) consists of a large-acceptance, non-focusing dipole magnet 
and a customized detector package.  For this experiment, the spectrometer was at an angle of 
99$^{\circ}$ w.r.t. the beam direction and 1.1~m from the target with a resulting angular acceptance of about 96~msr 
and a nominal momentum acceptance from 0.25~GeV/c to 0.9~GeV/c.  The detector 
package consisted of three planes of plastic scintillator 
segmented in the dispersive direction.  This unshielded system was able to run in Hall A 
with a luminosity of up to $10^{38}$~cm$^{-2}$s$^{-1}$. 

The non-focusing magnetic dipole, with a central field of 0.93~T, was used to bend the 
charged particles before they hit the detector planes. 
Timing, hit position, and energy deposited in the scintillators 
were all used to determine the incoming particle direction and momentum.
The detector achieved an angular resolution of 1.5~mrad in both the vertical 
and horizontal planes.  The timing resolution of the trigger plane 
scintillators was measured to be 0.5~nsec. 
This timing resolution along with path length correction translated to a momentum resolution of $\delta$p/p = 2.5\%.

\subsection*{Neutron Detector}

To detect recoiling neutrons, 88 plastic scintillator bars were placed directly 
behind the BigBite spectrometer at a distance of 6~m from the target. 
The detector had a width of 1~m, a height of 3~m, and a depth of 0.4~m.
A 5~cm thick lead wall was placed in front of the detector to 
block low-energy photons and most of the charged particles while allowing
most neutrons to pass.
A layer of 2~cm thick plastic scintillators was placed between the lead wall and the neutron detector
to identify any charged particles that managed to pass through the lead wall.
The detector covered a solid angle similar to that of BigBite. 

\begin {figure}[htb]
  \centering 
  %\vspace {30mm}         
  \includegraphics[width=\linewidth,angle=0]{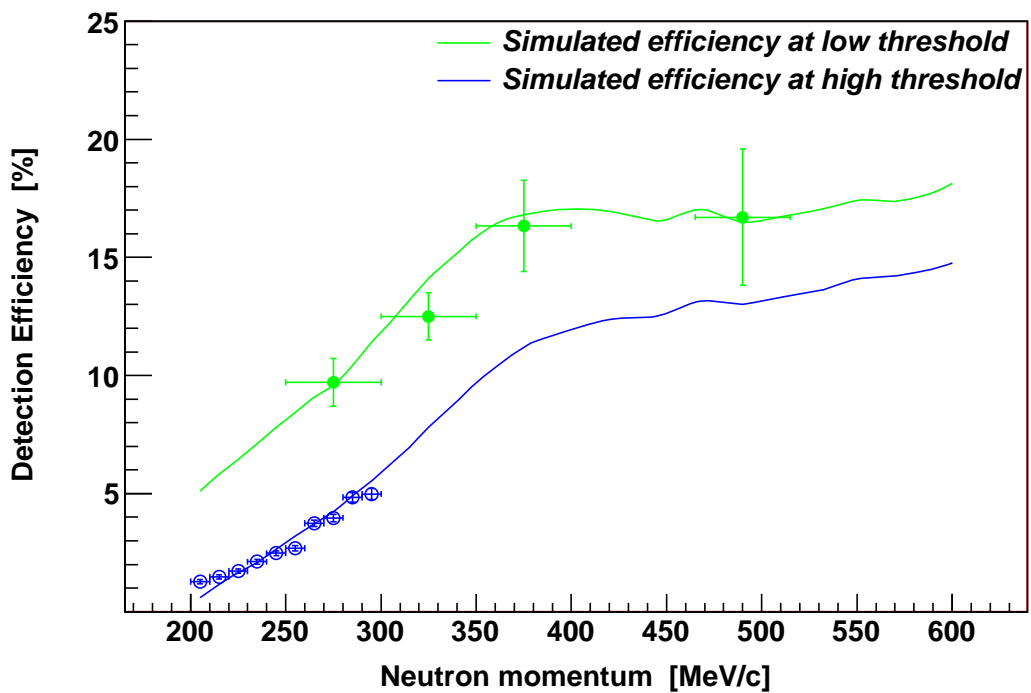}\\		      
  \caption{Shown is the neutron detection efficiency versus neutron momentum.  
The data are from the overdetermined quasi-elastic $^2$H$(e,e'p)n$ reaction, which was used
to make an effective neutron beam
while the curves are from a simulation.  The green curve is for a 4.627 GeV beam and a low detection threshold, while
the blue curve is for a 2.345 GeV beam with a high detection threshold.}
  \label{DetectionEff_2}
\end{figure}

The absolute probability for the detector to detect a neutron that 
originated at the target was determined using the $^2$H(e,e'p)n
reaction and checking the result against a 
simulation code that takes into account the attenuation 
of the neutron flux 
and the neutron detection efficiency of the plastic scintillators~\cite{Cecil:1979sz}.  
The results of this calibration are shown in Fig.~\ref{DetectionEff_2}.

\begin{table}[htp]
\label{SOM-Table-S1}
\begin{tabular}{l c c l c}
\hline Reaction	                               & Missing Momentum 	& Pair Fraction & Mean$\pm$SD	& Reference \\
					       & [MeV]			&		& [\%]       	&	\\
\hline $[^{12}$C(e,e'pp)/$^{12}$C(e,e'pn)$]$/2 & 0.41 - 0.51		& pp/pn		& 5.5 $\pm$ 1.5 & This Work \\
       $[^{12}$C(e,e'pp)/$^{12}$C(e,e'p)$]$/2  & 0.30 - 0.40		& pp/2N		& 3.7 $\pm$ 1   & \cite{Shneor:2007tu}	\\
       $[^{12}$C(e,e'pp)/$^{12}$C(e,e'p)$]$/2  & 0.40 - 0.50		& pp/2N		& 4.7 $\pm$ 1   & \cite{Shneor:2007tu} \\
       $[^{12}$C(e,e'pp)/$^{12}$C(e,e'p)$]$/2  & 0.50 - 0.60 		& pp/2N		& 4.7 $\pm$ 1   & \cite{Shneor:2007tu} \\
       $^{12}$C(e,e'pn)/$^{12}$C(e,e'p)        & 0.41 - 0.61		& pn/2N		& 96  $\pm$ 22  & This Work  \\
       $^{12}$C(p,ppn)/$^{12}$C(p,pp)          & 0.28 - 0.56 		& pn/2N		& 92  $\pm$ 18  & \cite{Piasetzky:2006ai} \\
\hline
\end{tabular}
\caption{Table of nucleon-nucleon pair fraction extracted along with the statistical error.
In the systematical error, the dominant contribution is the
correlated uncertainty of approximately 10\% of the mean for 
the  $^{12}$C(e,e'pn)/$^{12}$C(e,e'p) and $^{12}$C(e,e'pp)/$^{12}$C(e,e'p) results 
due to an acceptance correction which is limited by the knowledge of pair motion in the nucleus~\cite{Shneor:2007tu}.
For the ratio of $^{12}$C(e,e'pp)/$^{12}$C(e,e'pn) this systematic
effect cancels.}
\end{table}

\clearpage

%\bibliography{1156675bib}

\begin{thebibliography}{10}

\bibitem{Lapikas:1993}
L.~Lapikas, {\it Nucl. Phys.\/} {\bf A553}, 297 (1993).

\bibitem{Kelly:1996}
J.~Kelly, {\it Adv. Nucl. Phys.\/} {\bf 23}, 75 (1996).

\bibitem{Dickhoff:2004xx}
W.~H. Dickhoff, C.~Barbieri, {\it Prog. Part. Nucl. Phys.\/} {\bf 52}, 377
  (2004).

\bibitem{Egiyan:2003vg}
K.~S. Egiyan, {\it et~al.\/}, {\it Phys. Rev.\/} {\bf C68}, 014313 (2003).

\bibitem{Egiyan:2005hs}
K.~S. Egiyan, {\it et~al.\/}, {\it Phys. Rev. Lett.\/} {\bf 96}, 082501 (2006).

\bibitem{Niyazov:2003zr}
R.~A. Niyazov, {\it et~al.\/}, {\it Phys. Rev. Lett.\/} {\bf 92}, 052303
  (2004).

\bibitem{Benmokhtar:2004fs}
F.~Benmokhtar, {\it et~al.\/}, {\it Phys. Rev. Lett.\/} {\bf 94}, 082305
  (2005).

\bibitem{Aclander:1999fd}
J.~L.~S. Aclander, {\it et~al.\/}, {\it Phys. Lett.\/} {\bf B453}, 211 (1999).

\bibitem{Tang:2002ww}
A.~Tang, {\it et~al.\/}, {\it Phys. Rev. Lett.\/} {\bf 90}, 042301 (2003).

\bibitem{Malki:2000gh}
A.~Malki, {\it et~al.\/}, {\it Phys. Rev.\/} {\bf C65}, 015207 (2002).

\bibitem{Piasetzky:2006ai}
E.~Piasetzky, M.~Sargsian, L.~Frankfurt, M.~Strikman, J.~W. Watson, {\it Phys.
  Rev. Lett.\/} {\bf 97}, 162504 (2006).

\bibitem{Shneor:2007tu}
R.~Shneor, {\it et~al.\/}, {\it Phys. Rev. Lett.\/} {\bf 99}, 072501 (2007).

\bibitem{Frankfurt:1981mk}
L.~L. Frankfurt, M.~I. Strikman, {\it Phys. Rept.\/} {\bf 76}, 215 (1981).

\bibitem{Frankfurt:1993sp}
L.~L. Frankfurt, M.~I. Strikman, D.~B. Day, M.~Sargsian, {\it Phys. Rev.\/}
  {\bf C48}, 2451 (1993).

\bibitem{Alcorn:2004sb}
J.~Alcorn, {\it et~al.\/}, {\it Nucl. Instrum. Meth.\/} {\bf A522}, 294 (2004).

\bibitem{Mardor:1992sb}
I.~Mardor, Y.~Mardor, E.~Piasetzky, J.~Alster, M.~M. Sargsian, {\it Phys.
  Rev.\/} {\bf C46}, 761 (1992).

\bibitem{rocco2007}
R.~Schiavilla, R.~B. Wiringa, S.~C. Pieper, J.~Carlson, {\it Phys. Rev.
  Lett.\/} {\bf 98}, 132501 (2007).

\bibitem{Sargsian:2005ru}
M.~M. Sargsian, T.~V. Abrahamyan, M.~I. Strikman, L.~L. Frankfurt, {\it Phys.
  Rev.\/} {\bf C71}, 044615 (2005).

\bibitem{Lattimer:2004}
J.~M. Lattimer, M.~Prakash, {\it Science\/} {\bf 304}, 536 (2004).

\bibitem{Baym:1995}
G.~Baym, {\it Nucl. Phys.\/} {\bf A590}, 233 (1995).

\bibitem{Baym:2002}
G.~Baym, {\it Nucl. Phys.\/} {\bf A702}, 3 (2002).

\bibitem{Frick:2004th}
T.~Frick, H.~Muther, A.~Rios, A.~Polls, A.~Ramos, {\it Phys. Rev.\/} {\bf C71},
  014313 (2005).

\bibitem{Leemann:2001}
C.~W. Leeman, D.~R. Douglas, G.~A. Krafft, {\it Annu. Rev. Nucl. Part. Sci.\/}
  {\bf 51}, 413 (2001).

\bibitem{Cecil:1979sz}
R.~A. Cecil, B.~D. Anderson, R.~Madey, {\it Nucl. Instrum. Meth.\/} {\bf 161},
  439 (1979).

\end{thebibliography}
%
%\bibliographystyle{Science}

\begin{scilastnote}
\item
This work was supported by the Israel Science Foundation, the US-Israeli
Bi-national Scientific Foundation, the UK Engineering and Physical Sciences Research Council
and the Science \& Technology Facilities Council, 
the U.S. National Science Foundation, the U.S. Department of Energy grants DE-AC02-06CH11357,
DE-FG02-94ER40818, and U.S. DOE Contract No. DE-AC05-84150, Modification No. M175,
under which the Southeastern Universities Research
Association, Inc. operates the Thomas Jefferson National Accelerator Facility.  The raw data
from this experiment is archived in Jefferson Lab's mass storage silo in the directory /mss/halla/e01015/raw.
\end{scilastnote}

\end{document}